\documentstyle [epsf,12pt]{article}

\begin{document}
%
\renewcommand{\textfraction}{0}
\newcommand{\tcut}{t_{\mathrm{cut}}}
\makeatletter
\def\section{\@startsection{section}{3}{\z@}{-3.25ex plus
 -1ex minus-.2ex}{1.5ex plus.2ex}{\reset@font\normalsize\bf}}
\makeatother

\title{\Large Charm lifetime\thanks{Talk presented at the 6th.\ International
Symposium on Heavy Flavour Physics (Pisa, June 1995).}}
\author{\large S. Malvezzi \\
               I.N.F.N., via Celoria 16, 20133 Milano}
\date{June 1995}
\maketitle


\begin{abstract} \noindent
A review of the charmed meson and baryon lifetimes is presented. Our knowledge
of charmed particle lifetimes has greatly improved over the past two years, a
crucial r\^ole having been played by the E687 experiment at Fermilab, which has
almost quadrupled the samples of $D$ mesons. The lifetime ratios
$\tau(D^+)/\tau(D^0)$ and $\tau(D_s^+)/\tau(D^0)$ are now known with an
accuracy of 1.7\% and 3.7\% respectively. In the baryon sector the statistics
is still limited, but the experimental results on $\Lambda_c^+$, $\Xi_c^0$ and
$\Xi_c^+$ exhibit a clear pattern of lifetime hierarchy, as expected from
simple theoretical arguments. The first measurement of $\tau(\Omega_c^0)$ from
E687 is also presented to complete the charmed baryon lifetime picture. The
more accurate experimental scenario can provide information on non-perturbative
QCD effects and the hadronic matrix elements.
\end{abstract}


\begin{flushleft}
\end{flushleft}


\normalsize
\section{The issue of charm lifetimes}

Heavy flavour hadrons consist of a heavy quark $Q$ and a cloud of light quarks
and gluons. Since the early days of QCD it has been known that, in the limit of
infinite mass $m_Q$, the total decay rates and lifetimes of weakly decaying
particles containing the same flavour $Q$ should be identical and equal to that
of a ``free'' heavy decay, the effect of the bound state being neglected in
calculating the decay probability. This na\"{\i}ve intuition is contradicted by
the experimental measurements of the charm family lifetimes, which manifest
wide disparity between mesons and baryons (roughly a factor of 10 between
$\Xi_c^0$ and $D^+$) and a significant difference between $D^+$ and $D^0$
(roughly a factor of 2).

This picture indicates that preasymptotic corrections, due to interactions with
soft degrees of freedom in the light cloud, cannot thus be disregarded at the
charm-mass scale; the inclusion of bound-state effects generates
non-perturbative power corrections in $1/m_c$. Moreover, the presently
available precise measurements of the ratio
$R=\tau(D^+)/\tau(D^0)=2.547\pm0.043$ and of the inclusive branching ratios
$D^+\to eX=(17.2\pm1.9)\%$ (PDG94) and $D^0\to eX=(6.97\pm0.18\pm0.30)\%$
(CLEO) lead to an estimate of almost equal semileptonic decay widths for the
$D^0$ and $D^+$:
$$
 {\Gamma(D^0 \to eX) \over \Gamma(D^+ \to eX)}
 = {Br(D^0 \to eX) \over Br(D^+ \to eX)} \times {\tau(D^+) \over \tau(D^0)}
 = 1.03 \pm 0.12,
$$
revealing that the extra rate in the total decay widths has to be looked for in
the hadronic sector and that QCD corrections have to be taken into account. It
is worthwhile pointing out that the lifetime differences in charmed mesons
should be understood at the level of two-body decays; recent results on Dalitz
plot analyses indicate that the high-multiplicity decays are dominated by
resonant substructures, the only exception being $D^+\to K^-\pi^+\pi^+$, which,
alone among the $D$ decays already analyzed, cannot be described without a
large three-body non-resonant component.

\section{Theory versus experiment}

Preasymptotic effects in the meson sector enter in the form of $W\!$-exchange
and $W\!$-annihilation diagrams, correcting the na\"{\i}ve spectator scenario;
additional contributions come from Pauli interference of the decay quark and
spectator quarks in the $D^+$ only, lengthening the lifetime of the $D^+$ with
respect to the $D^0$. In contrast to the main spectator decay, the evaluation
of both these effects requires a knowledge of the $D$ wave function.
$W\!$-exchange and annihilation are expected to be inhibited by helicity
conservation although soft gluon emission would attenuate the suppression.
Quantitative estimates of both gluon enhancement and non-spectator
contributions are difficult and subject to uncertainty. Anyway, the inclusion
of these effects can account for the experimentally observed hierarchy:
$$
  \tau(D^+)>\tau(D^0) \approx \tau(D_s^+).
$$
The $D_s\to3\pi$ decay represents a unique window to evaluate the annihilation
rate. A Dalitz-plot analysis of this channel allows the separation of
components that only proceed through annihilation: namely, the three-body
non-resonant and, in the absence of hostile final-state interactions, the
$\rho^0\pi^+$ contributions.

Analogously, the baryon sector manifests preasymptotic effects in the form of
quark interference and $W\!$-exchange. The essential difference from the meson
case, however, lies in the fact that $W\!$-exchange among valence quarks of the
baryon is neither helicity nor colour suppressed. Moreover, soft-gluon
radiation, the crucial question in meson decays, is a mere correction to the
pure valence-quark process. Baryons are thus an ideal laboratory for studying
preasymptotic effects; they also offer the advantage of access to four states
whose different quark structure implies different combinations of
$W\!$-exchange and interference of different strengths. The relative size of
these effects leads to different predictions~\cite{Teorie}:
\begin{eqnarray*}
 \tau(\Omega_c^0)\approx\tau(\Xi_c^0)<\tau(\Lambda_c^+)<\tau(\Xi_c^+),
 & \quad &
 \tau(\Omega_c^0)<\tau(\Xi_c^0)<\tau(\Lambda_c^+)\approx\tau(\Xi_c^+)
 \\
 &\hbox to0pt{\hss\centerline{or\qquad
 $\tau(\Omega_c^0)\le\tau(\Xi_c^0)<\tau(\Lambda_c^+)<\tau(\Xi_c^+).$}\hss}&
\end{eqnarray*}
The hierarchy-pattern ends are due to the absence of $W\!$-exchange for the
$\Xi_c^+$ and to large positive interference among $s$ quarks in the
$\Omega_c^0$. Previous analyses have recently been revisited and
improved~\cite{Blok}, concluding that the $\Omega_c^0$ could be either the
shortest or longest living among the weakly decaying baryons depending on the
strength of the spin-spin interaction and the value of the $D$ decay constant.
An evaluation of the different effects can also be found in ref.~\cite{Gupta}.

In the meson sector the ratio $\tau(D^+)/\tau(D^0)=2.547\pm0.043$ is measured
to 1.7\%; the destructive interference invoked between $D^+$ external and
internal spectator diagrams can account for this factor 2 difference.
The \hbox{$D_s^+$-$D^0$} lifetime ratio, $\tau(D_s^+)/\tau(D^0)=1.125\pm0.042$,
is also now measured to good accuracy (3.7\%) and turns out to deviate
from the predicted value of unity at the $3\sigma$ level, thus raising the
question of whether theoretical estimates can accomodate this difference.
Precise information on this ratio provides a sensitive gauge of the impact of
weak annihilation in charm decays and of the weight of $SU(3)_f$ breaking.

In the baryon sector different hierarchy patterns are predicted; they reflect
intimate features of hadronic structure and are to a large extent model
independent; {\it i.e.}, they do not require a knowledge of the $B$ baryon
wave-function. Thus, fairly accurate measurements would permit disentangling
and determining the individual spectator and non-spectator effects. Once
established, the observed hierarchy would imply relations between hadronic
matrix elements of the operators involved, unique information otherwise
inaccessible. Unfortunately, in the baryon case we do not have the powerful
tool of factorization (applicable to mesons). From this point of view,
experimental information is necessary to probe our understanding of weak-decay
dynamics at preasymptotic scales.

It should not be forgotten that the QCD-based inclusive approach will never
become fully quantitative in the charmed family; the $c$-quark lies below or at
the border of the domain where the heavy-mass expansion may be useful, which
makes the task challenging. The enhanced r\^ole of preasymptotic effects leads
to the necessity of carefully considering all operators to be included in the
matrix elements: some updated analyses are now being produced~\cite{Blok,Bigi}.
The theory issue then concerns the size of the lifetime differences and
absolute lifetime predictions. The measurement of the doubly strange
$\Omega_c^0$ baryon lifetime has turned out to be crucial; here strong spin
terms, as in the mesons, are present and the conjecture of an unknown mechanism
(distinguishing between mesons and antitriplet baryons) somehow connected with
the spin terms can be tested.

\section{The experimental scenario}

The experimental scenario is chronologically portrayed in tables
\ref{tab:Dlife} and \ref{tab:LCOlife}. The $D$ meson statistics has been
\begin{table}[hbt]
\caption{$D^+$, $D^0$, $D^+_s$ lifetime measurements.}
\centering
\begin{tabular}{||l|c|c|c|r||}\hline
Meson
& Experiment & Year & Beam, reaction                             & Events\cr
\hline
$D^+$
& E687       & 1993 & $\gamma$ Be, $D^+ \to K^-\pi^+\pi^+$       &  9189 \cr
& NA14       & 1990 & $\gamma$, $D^+ \to K^-\pi^+\pi^+$          &   200 \cr
& ACCMOR     & 1990 & $\pi^-$ Cu 230 GeV                         &   317 \cr
& ARGUS      & 1988 & $e^+e^-$ 10 GeV                            &   363 \cr
& E691       & 1988 & Photoproduction                            &  2992 \cr
& LEBC-EHS   & 1987 & $\pi^- p$ and $pp$                         &   149 \cr
& CLEO       & 1987 & $e^+e^-$ 10 GeV                            &   247 \cr
\hline
$D^0$
& E687       & 1993 & $\gamma$ Be, $D^0 \to K^-\pi^+, K^-3\pi$   & 16730 \cr
& NA14       & 1990 & $\gamma$, $D^0 \to K^-\pi^+, K^-3\pi^-$    &   890 \cr
& ACCMOR     & 1990 & $\pi^-$ Cu 230 GeV                         &   641 \cr
& ARGUS      & 1988 & $e^+e^-$ 10 GeV                            &   776 \cr
& E691       & 1988 & Photoproduction                            &  4212 \cr
& LEBC-EHS   & 1987 & $\pi^- p$ and $pp$                         &   145 \cr
& CLEO       & 1987 & $e^+e^-$ 10 GeV                            &   317 \cr
\hline
$D^+_s$
& E687       & 1993 & $\gamma$ Be, $D_s^+ \to \phi\pi^+$         &   900 \cr
& NA14       & 1990 & $\gamma$, $D_s^+ \to \phi\pi^+$            &    15 \cr
& ACCMOR     & 1990 & $\pi^-$ Cu 230 GeV                         &    54 \cr
& HRS        & 1989 & $e^+e^-$ 29 GeV                            &    18 \cr
& ARGUS      & 1988 & $e^+e^-$ 10 GeV                            &   144 \cr
& E691       & 1988 & Photoproduction                            &   228 \cr
& CLEO       & 1987 & $e^+e^-$ 10 GeV                            &   141 \cr
\hline
\end{tabular}
\label{tab:Dlife}
\end{table}
\begin{table}[hbt]
\caption{$\Lambda_c^+$, $\Xi_c^+$, $\Xi_c^0$ and $\Omega_c^0$ lifetime
measurements.}
\centering
\begin{tabular}{||l|c|c|c|r||}\hline
Baryon
& Experiment & Year & Beam, reaction                                & Events
\cr
\hline
$\Lambda_c^+$
& E687       & 1993 & $\gamma$ Be, $\Lambda_c^+\to pK^-\pi^+$       & 1340
\cr
& NA14       & 1990 & $\gamma$, $\Lambda_c^+\to pK^-\pi^+$          &   29
\cr
& ACCMOR     & 1989 & p$K^-\pi^+$ + c.c.                            &  101
\cr
& E691       & 1988 & Photoproduction                               &   97
\cr
& LEBC 	     & 1988 & $\pi^-p$ and $pp$                             &    9
\cr
\hline
$\Xi_c^+$
& WA89  & 1994 & $\Sigma^-$ Cu-C, $\Xi_c^+\to\Lambda K^-\pi^+\pi^+$ &   20
\cr
& E687         & 1993 & $\gamma$ Be, $\Xi_c^+\to\Xi^-\pi^+\pi^+$    &   30
\cr
& ACCMOR       & 1989 & $\pi^-(K^-)$ Cu 230 GeV                     &    6
\cr
& E400         & 1987 & nA $\sim$ 600 GeV                           &  102
\cr
& Biagi {\it et al.} & 1985 & Hyperon beam                          &   53
\cr
\hline
$\Xi_c^0$
& E687         & 1993 & $\gamma$ Be, $\Xi_c^0\to\Xi^-\pi^+$         &   42
\cr
& ACCMOR       & 1990 & $\pi^-(K^-)$ Cu 230 GeV                     &    4
\cr
\hline
$\Omega_c^0$
& E687         & 1995 & $\gamma$ Be, $\Omega_c^0\to\Sigma^+K^-K^-\pi^+$ & 43
\cr
& WA89         & 1995 & $\Sigma^-$ Cu-C (expected soon)             & \cr
\hline
\end{tabular}
\label{tab:LCOlife}
\end{table}
steadily increasing over the years with, basically, two milestones: the E691
and E687 fixed-target photoproduction experiments at Fermilab. In the baryon
field the statistics is still limited and measurements are affected by large
errors, but data are now becoming available and we are also benefitting from
the first result on $\Omega_c^0$ lifetime from E687, completing the four-baryon
scenario. At this conference a very preliminary estimate of $\tau(\Omega_c^0)$
from WA89 has been presented; a more reliable measurement will hopefully be
available soon. The experimental scenario is dominated by E687, which alone has
measured all the charmed hadron lifetimes; the results are thus checked as
internally consistent and the relative ratios, characterizing the hierarchy
patterns, are to a large extent unbiased by systematic effects. Details of the
experiment can be found in ref.~\cite{Exp:E687Desc} and the general methodology
applied in refs.~\cite{lifetime}. In tables \ref{tab:Dcomp} and
\begin{table}[hbt]
\caption{Lifetimes of the $D^+$, $D^0$, $D_s^+$ (ps).}
\centering
\begin{tabular}{||l|c|c|c|c||}\hline
        & E687                         & PDG94       & Accuracy & Accuracy \cr
        &                              &             & (E687)   & (PDG94)  \cr
\hline
$D^+$   & $1.048 \pm 0.015 \pm 0.011$  & $1.057 \pm 0.015$ & 1.8\% & 1.4\% \cr
$D^0$   & $0.413 \pm 0.004 \pm 0.003$  & $0.415 \pm 0.004$ & 1.2\% & 1.0\% \cr
$D_s^+$ & $0.475 \pm 0.020 \pm 0.007$  & $0.467 \pm 0.017$ & 4.5\% & 3.6\% \cr
\hline
\end{tabular}
\label{tab:Dcomp}
\end{table}
\ref{tab:LCOcomp} the Particle Data Group 1994 meson and baryon lifetimes are
\begin{table}[hbt]
\caption{Lifetimes of the $\Lambda_c^+$, $\Xi_c^0$, $\Xi_c^+$,
$\Omega_c^0$~(ps).}
\centering
\def\Z{\hphantom{0}}
\def\Y{\hphantom{\scriptstyle0}}
\def\X#1{\rlap{$\;\;{}#1$}\hphantom{\null\pm0.000}}
\begin{tabular}{||l|c|c|c|c||}\hline
& E687                               & PDG94    & Accuracy & Accuracy  \cr
&                                    &          & (E687)   & (PDG94)   \cr
\hline
$\Lambda_c^+$
& $0.215\pm0.016\pm0.008$                    & $0.200  \; {}^{+0.011}_{-0.010}$
& \Z8.3\%    & \Z5.5\%  \cr
$\Xi_c^0$
& $0.101  \X{^{+0.025}_{-0.017}} \pm 0.005$  & $0.098  \; {}^{+0.023}_{-0.015}$
&  25.2\%   &  23.4\% \cr
$\Xi_c^+$
& $0.41\Z \X{^{+0.11}_{-0.08}}   \pm 0.02\Z$ & $0.35\Z \; {}^{+0.07}_{-0.04}\Y$
&  27.2\%   &  20.0\% \cr
$\Omega_c^0$
& $0.089  \X{^{+0.027}_{-0.020}} \pm 0.028$  &
&  43.7\%   &        \cr
\hline
\end{tabular}
\label{tab:LCOcomp}
\end{table}
reported and compared with the E687 final results, fractional errors are also
quoted. The ratio between $D^+$ and $D^0$ lifetimes is thus known to within
1.7\% and the ratio between $D_s^+$ and $D^0$ lifetimes with 3.7\% accuracy:
$$
  {\tau(D^+)   \over \tau(D^0)} = 2.547 \pm 0.043 , \qquad
  {\tau(D_s^+) \over \tau(D^0)} = 1.125 \pm 0.042 \;.
$$
The lifetime ratios between baryons can also be extracted:
$$
  {\tau(\Xi^+_c) \over \tau(\Lambda_c^+)} = 1.75 \pm 0.36 , \qquad
  {\tau(\Xi^0_c) \over \tau(\Lambda_c^+)} = 0.49 \pm 0.12 \;.
$$
Since PDG94 a new measurement of $\tau(\Xi^+_c)$ by WA89 has been made
available: $\tau(\Xi^+_c)=0.32\,^{+0.08}_{-0.06}\pm0.05$~ps, in very good
agreement with the world average.


\section*{The first measurement of the $\Omega_c^0$ lifetime}

\begin{figure}[hbt]
\centering
\makebox[8cm]{\epsfxsize=8cm \epsfbox{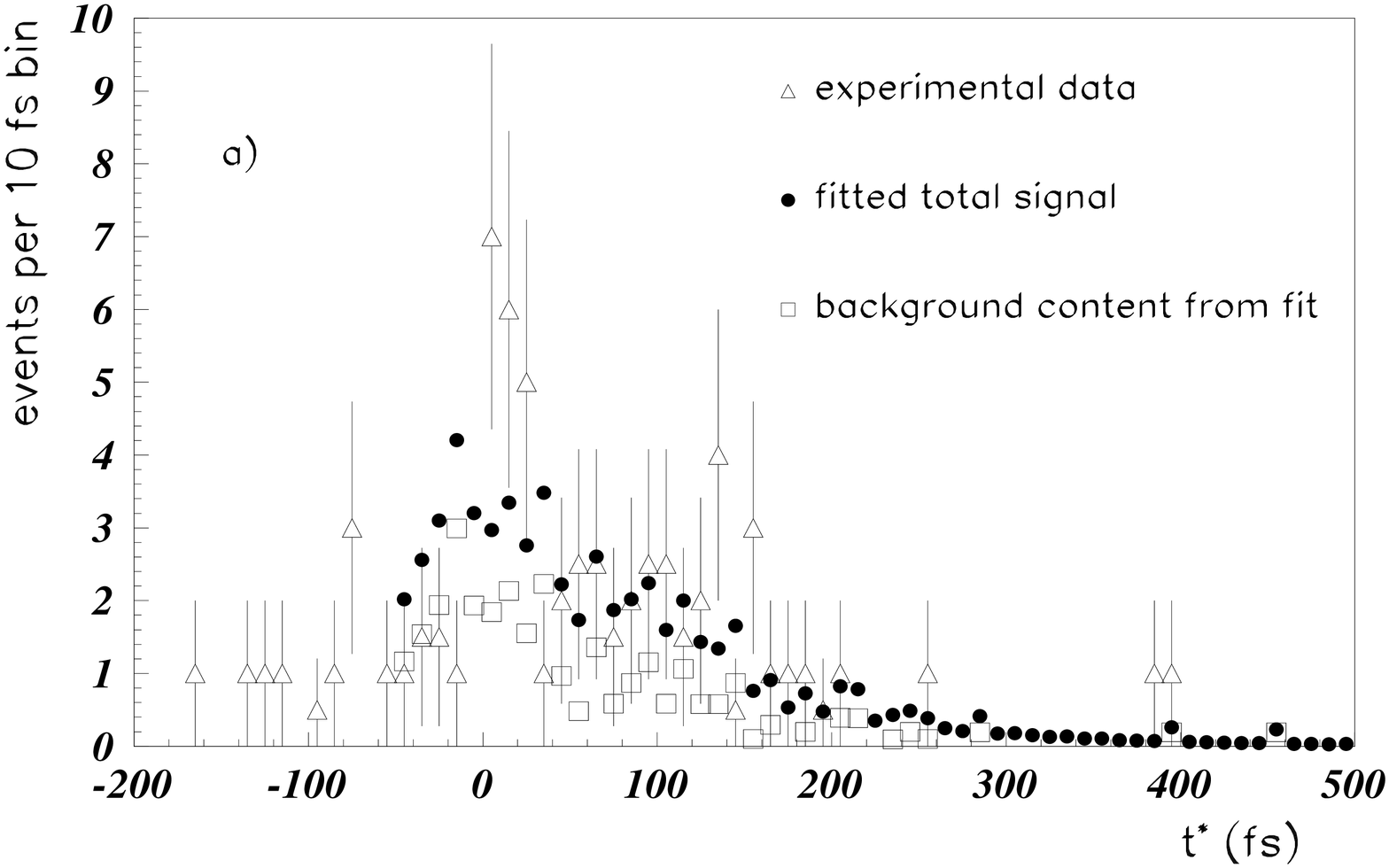}}
\hspace{0.1cm}
\makebox[6cm]{\epsfxsize=6cm \epsfbox{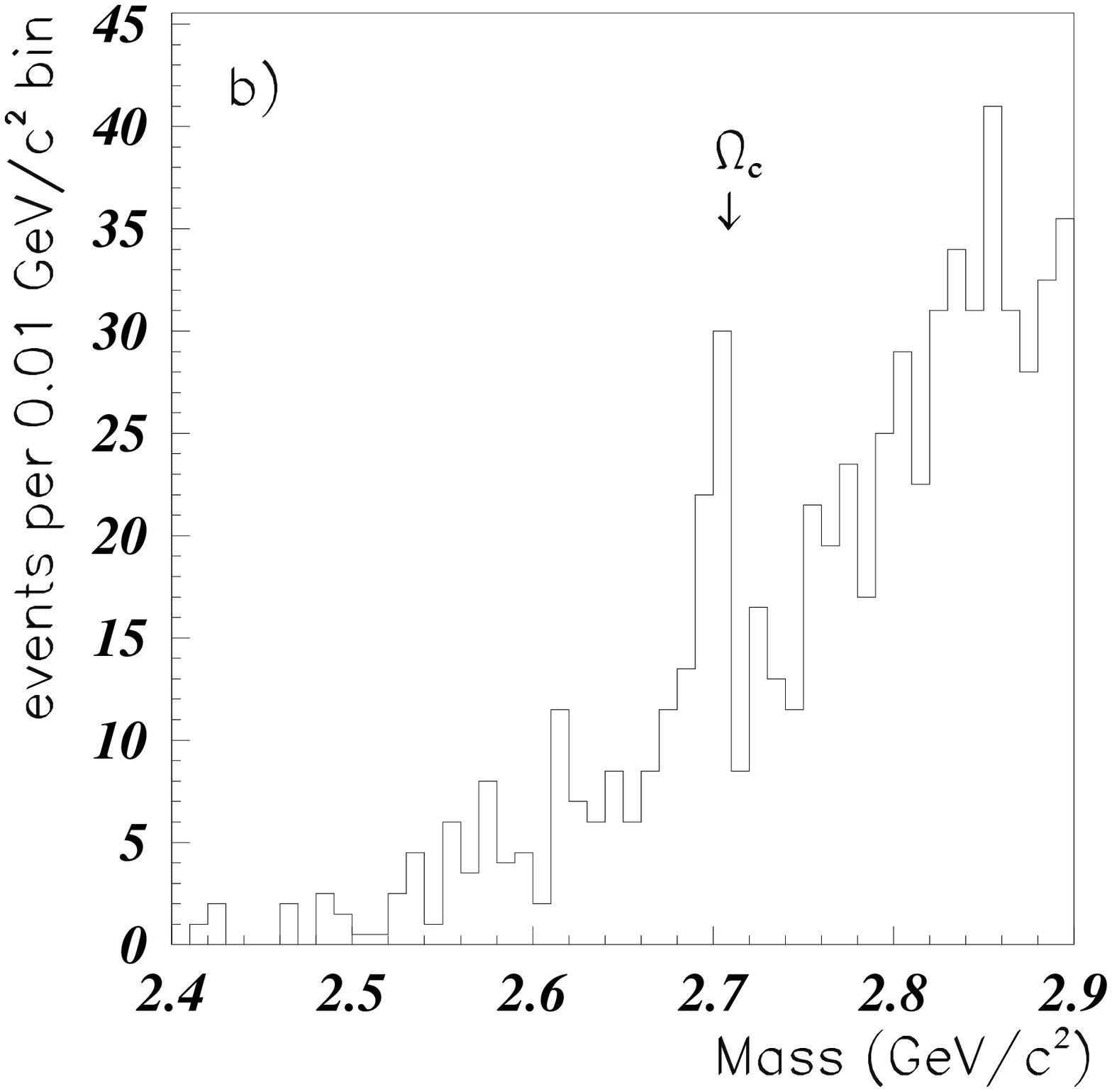}}
\caption {a) Proper-time evolution and b) mass distribution of the $\Omega_c^0$
events.}
\label{fg:masse}
\end{figure}

Among the baryons, the most important experimental result is the first
measurement of the $\Omega_c^0$ lifetime, performed by the E687 collaboration
in the channel $\Omega_c^0\to\Sigma^+K^-K^-\pi^+$; evidence of this decay and
details of the relative analysis have already been published by
E687~\cite{omegamass}. The raw proper-time distribution in Fig.~\ref{fg:masse}a
indicates that the lifetime is quite short, comparable to the spectrometer time
resolution, which is predicted by Monte Carlo to be $\sim0.070-0.080$~ps. To
successfully measure this short lifetime, very accurate account of proper-time
distribution smearing due to finite vertex resolution is necessary. Since
smearing will cause some true signal events to appear to have negative
proper-time, such events have to be included in the fit. To study the stability
of the fit, to begin with all the events are fitted and then only those with
proper time greater than a cut value, $\tcut^*$. To quote the final lifetime
value, the collaboration adopts a $\tcut^*>-0.050$~ps, as a compromise between
a good signal-to-noise ratio and the largest possible statistics. The chosen
sample also gives the best fit probability based on the Kolmogorov-Smirnov
test.

\begin{figure}[hbt]
\centering
\makebox[6.2cm]{\epsfxsize=6.4cm \epsfbox{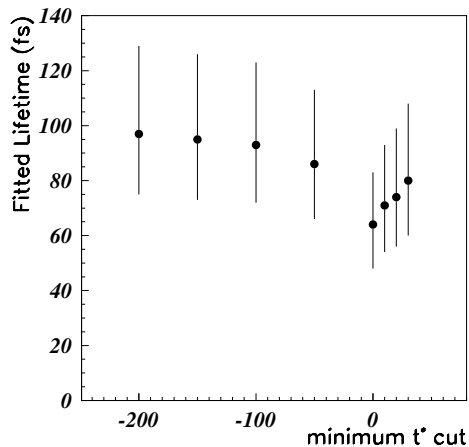}}
\caption{Lifetime of the $\Omega^0_c$ for different minimum $t^*$ cuts.}
\label{fg:varit}
\end{figure}

Fig.~\ref{fg:masse}b shows the mass distribution of events considered in the
final fit. The results of the fits to all events and to events with proper
times greater than a cut value, defined as $\tcut^*$, are shown in
Fig.~\ref{fg:varit}; within the statistical errors returned by the fit, all the
fitted lifetime values are consistent. The large relative variation of the
fitted lifetimes obtained by using different $\tcut^*$ values indicates a
possible systematic uncertainty, for which an upper limit of 0.0129~ps is
estimated. Additional systematic contributions may come from the sideband
locations, background lifetime, Monte Carlo simulation uncertainty and
absorption of secondaries. All these effects are estimated and a final value
for the $\Omega_c^0$ lifetime is quoted as
$$
 \tau(\Omega_c^0) = 0.089\,^{+0.027}_{-0.020}
                    \;\mbox{(stat.)} \pm 0.028 \;\mbox{(syst.) ps}.
$$

This result establishes the first measured full lifetime-hierarchy pattern:
$$
  \tau(\Omega_c^0) \le \tau(\Xi_c^0) < \tau(\Lambda_c^+) < \tau(\Xi_c^+)
  < \tau(D^0) < \tau(D^+_s) < \tau(D^+).
$$


\section{Conclusions}

The charmed-hadron lifetime field is now rather mature; in the meson sector the
measurements are very precise (1--2\%) and evidence for $\tau(D^+_s)>\tau(D^0)$
is provided at the $3\sigma$ level. These results pose severe new constraints
on theoretical models describing charm decay, their precision probably
exceeding the ability to compute them. In the baryon sector a significant
improvement over previous $\Lambda_c^+$, $\Xi_c^+$ and $\Xi_c^0$ lifetime
measurements has been reached in the last two years. The first measurement of
the $\Omega_c^0$ lifetime has also been performed by E687, making available a
full lifetime-hierarchy pattern. More data on baryons are certainly needed and
the continuous experimental effort to deal with $\tau\le10^{-13}$~s is
worthwhile. Progress from E687/E831, WA89 and EXCHARM is expected soon.

On the theory side, efforts to develop systematic and self-consistent
theoretical treatments of the inclusive weak decays of Heavy Flavour hadrons
have been made. Hopefully, the cooperation between different {\em second
generation technology\/} analyses (QCD sum rules, HQET, $1/m_Q$ expansion,
lattice QCD {\it etc.\/}) will be fruitful. Charm lifetime studies are
providing very useful information on perturbative and non-perturbative QCD
effects, necessary to understand the phenomenology of the lightest of the heavy
quarks; results on the charm family may therefore be important for predictions
in the beauty sector and, hopefully, help illuminate the light quark world.



\end{document}